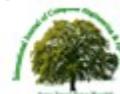

# A NOVEL TERM WEIGHING SCHEME TOWARDS EFFICIENT CRAWL OF TEXTUAL DATABASES

**Sonali Gupta[1], Komal Kumar Bhatia [2]**

[1]*Department of Computer Engineering YMCA University of Science & Technology Faridabad, India*

[2]*Department of Computer Engineering YMCA University of Science & Technology Faridabad, India*

**ABSTRACT:**

The Hidden Web is the vast repository of informational databases available only through search form interfaces, accessible by therein typing a set of keywords in the search forms. Typically, a Hidden Web crawler is employed to autonomously discover and download pages from the Hidden Web. Traditional hidden web crawlers do not provide the search engines with an optimal search experience because of the excessive number of search requests posed through the form interface so as to exhaustively crawl and retrieve the contents of the target hidden web database. Here in our work, we provide a framework to investigate the problem of *optimal search* and curtail it by proposing an effective query term selection approach based on the *frequency & distribution* of terms in the document database. The paper focuses on developing a term-weighing scheme called *VarDF (acronym for variable document frequency)* that can ease the identification of optimal terms to be used as queries on the interface for maximizing the achieved coverage of the crawler which in turn will facilitate the search engine to have a diversified and expanded index. We experimentally evaluate the effectiveness of our approach on a manually created database of documents in the area of Information Retrieval.

## [I] INTRODUCTION

The Hidden Web refers to a huge portion of the World Wide Web (WWW) that holds numerous freely accessible Web databases, hidden behind search form interfaces which can only be accessed through dynamic web pages that are generated in response to the user queries issued at the search form interface. [9, 11]. A Hidden Web database depending on the type of its contents can be categorized as structured or unstructured. The structured databases provide multi-attribute search interfaces that have multiple query boxes pertaining to different aspects of the content. [Figure-1] is an example of such a form interface.

![Author / Title / ISBN(s) / Publisher / Subject form interface]

Fig: 1. A multi-attribute search form interface for an online book store

The unstructured databases however, usually contain plain-text documents which are not well structured and thus provide a simple keyword-based search interface having an input control (text type) for the user to type in a list of keywords for filling it. [Figure-2] shows an example of such a keyword based search interface.

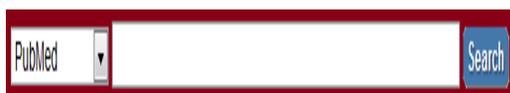

**Fig: 2. Keyword-based Search Interface**

The task of any Hidden web crawler involves pre-computing the most relevant form submissions for all interesting HTML (Hypertext Mark Up Language) forms by the crawler module, generating the resulting (Uniform Resource Locators )URLs offline and adding the obtained HTML pages into the search engines index[2,3,5]. This solution raises the issue of automatically selecting the valid input values to be submitted to the search inputs of the different forms.

Filling these simple search boxes with different relevant terms may return the same set of documents from the database. Posing all of them would not be a good retrieval strategy as it yields the same set over and over again. Intuitively, posing some optimal set of terms that can retrieve a diversified document set covering the entire Hidden database is more desirable than posing the entire set of terms that retrieves and presents the same set of documents repeatedly. Thus, it becomes necessary not only to rank the terms but also select an optimal set from the ranked terms. Our work focuses on how to formulate appropriate query terms for text box based simple search form interfaces that can yield maximum number of documents from the database by posing a minimum number of such terms sequentially on the interface.

## [II] STATE OF THE ART

It's a long standing challenge to develop an effective retrieval model that can ease the users search by providing an optimal crawl of the Hidden Web resources. Certain proposals [2, 3, 4, 5, 6, 7] have been made to extract the content hidden behind the search forms.

The work in [2] proposed HiWE, Hidden Web Exposer, a task-specific hidden-Web crawler, the main focus of this work is to learn Hidden-Web query interfaces. Their strategy aims to extract the labels of the form by rendering the pages. The crawler makes several filling attempts by testing various combinations of the values for HTML search forms available at the moment of crawl. Liddle et al. [3] performs a comprehensive study on obtaining valuable information from the web forms, but do not include a crawler to fetch them. Barbosa and Freire [4] experimentally evaluate methods for building multi-keyword queries that can return a large fraction of a document collection. Ntoulas et al. [5] differs from the previous studies, that, it provides a theoretical framework for analyzing the process of generating queries for a database problem of Hidden Web crawling. Gravano [6] have developed an automatic query-based technique to retrieve documents for extracting user-defined relations from large text databases, which can be adapted to databases from various domains or types (multi-attribute databases) with minimal human effort.

Gupta and Bhatia [9] developed a domain Oriented Hidden Web crawler, HiCrawl targeted to crawl the sites in the 'Medical' domain. The system is based on the use of classification hierarchies that might have either been manually or automatically constructed for choosing the right query term to be submitted to any search form interface that has been designed to accept keywords or terms as input to it. Whereas the work in [7] present a formal framework that regards the crawler as an agent that perceives its current state and the last execution action as input so as to output the next optimal action to be taken. It uses adaptive approach to reinforcement learning for deep web crawling where the action causes the agent (crawler) to transit from the current state to the next or successor state.

Most of these methods generate candidate query keywords either by using a manually or semi automatically created database of values or by using only the frequency statistics derived from the records obtained by previously issued queries without considering the structural properties of the HTML documents. Our approach of crawling differs from other adaptive approaches (based on results obtained from previous queries) by taking into consideration the position as well as the distribution of the terms in the document space, unlike others. The measure proposed for use in our approach to weigh and rank terms has been termed as the variable document frequency (Vardf) and is based on the fact that the vocabulary set changes and new structural properties of documents come up, as and by different documents are retrieved from the database. Thus, certain features should be re-estimated after any term has been selected and used as the query term.

## [III] PROBLEM FORMULATION

A textual database is a site that mainly contains plain-text documents. We assume D= {$d_1$, $d_2$, $d_3$, ……..$d_m$ } as the document database and Q={$t_1$, $t_2$, $t_3$,……..$t_n$} as the vocabulary set . If single term queries are considered as input to the text box, every potential query $q_i$ is a term from the term space. So the vocabulary set forms the set Q, the Query term space. Also, this vocabulary set has been constructed after removing the various stop words like Pronouns (we, she, his, her, its), verbs (is, were, are, did), Anaphors(this, that), Honorees (Dr, Ms., Mrs.) , conjunctions(but, and, or, so), prepositions (in, on , near, far, of, for, by, to , with) etc. from the term space.

Each query term $q_i$ is expected to convey some information through the retrieval of the documents from the textual database, thus at any instance a query typically is $q_i$

where $q_i$ ε Q. Suppose $2^Q$ denotes the power set of Q with $|2^Q| = 2^m$:

$2^Q$={$Q_i$ | $Q_i$ subset of Q}. The problem is to choose an optimal subset $Q_{opt}$ so as to maximize the achieved coverage of the crawler. The choice of the optimal subset depends on

- Size of the subset: There exists no subset $Q_i$ of $Q_{opt}$ which yields the same coverage.

*Cover ($Q_{opt}$) ≥ Cover ($Q_i$)* ∀$Q_i$, where $Q_i$ is a subset of $Q_{opt}$

- Coverage gain of the crawler: The coverage gain between the set Q and $Q_{opt}$ is maximum,

$$Q_{opt}= \max_{Q_i \in 2|Q|} \left\{ \frac{Cover(Q)-Cover(Q_i)}{Cover(Q)} \right\}$$

Where Cover(x) is a function measuring the number of fresh documents that have not been retrieved by issuing previous queries. It can be reasonably assumed that two different query terms may return either the same or slightly different or totally different sets of documents. Intuitively, the obtained coverage can be maximized if none of the query terms retrieve the same set of documents from the database. Achieving entirely independent and different sets of documents is practically infeasible.

This triggers the action of checking the coverage against each term in the query space. But this does not seem suitable for the huge size of the databases in the enormously sized Hidden Web. The problem can thus be simplified by choosing the most effective query term $q_i$ from Q such that the coverage gain between Q and Q-$\{q_i\}$ is maximized:

$$q_{opt} = \max_{qi \in Q} \left\{ \frac{Cover(Q) - Cover(Q - qi)}{Cover(Q)} \right\}$$

Once the effective $q_i$ is selected, $Q_{opt}$ could be approximated by iteratively adding and choosing local effective terms from Q. Thus, the $Q_{opt}$ will be generated by greedily selecting effective terms. The algorithm presented in the [Figure-3] justifies the use of $Q_{opt}$ in crawling the hidden Web.

HIDDEN WEB CRAWLING ALGORITHM

Input   :  - search form interface           ; The optimal
query set Q    opt  : the size N of the database      &
its domain

Ouptut   :  Documents from the Hidden Web
Database

Method    :  Analyze the form interface for valid
inputs   , decide the term to be issued as query
on the interface  , process the form interface
by issuing the query    , retrieve the results      ,
add to search engines index           .

1)  Download the search form interface              F.
2)  If keyword    _based    (f)= TRUE     analyze the
    input control            ;
3)  If input    _control    0= TEXTBOX
    While  { resources   (crawler ) && Q_opt != EMPTY
                        && coverage   (db )!= N}
        for all    (q_i ∈ Q_opt )
        SearchDatabase     (q_i) ;
        Download  (doc (q_i)) ;
        Cover  (q_i)= # of unique documents
        containing the term q              : ;
        Coverage   (db )= Coverage   (db ) + Cover  (q_i) ;

Fig: 3. Proposed crawl mechanism for Unstructured Databases in the Hidden Web.

## [IV] EFFECTIVE TERM SELECTION APPROACH

The contents of a Hidden web database usually focus on a single topic or domain. Thus, the same term in all the different documents capture the same interpretation rather than reflecting several different aspects. For example, for a database that belongs to FOOD domain, posing the term "java", will return a set of documents about coffee giving the complete picture of the database and no point lies in posing the term "programming" or "language" as input to the search interface.

So, it is important to pose such an optimal diversified set of terms on the search interface that can cover a different set of documents with every different term and yields the maximum number of records by posing such an optimal set of effective terms. Thus, the key challenge becomes identifying the terms that are semantically meaningful for a given Hidden Web database. To achieve maximum coverage, not only the high frequency terms but also the terms with low hit numbers be used should be used for queries. So the optimal set should contain a diversified set of terms which can benefit both the over-specified and underspecified terms without being diverted from the main goal of achieving maximum or exhaustive coverage of the database.

## 4.1. The VarDF Weighing Scheme

The same document in the database may be retrieved by invoking several different terms on the search interface. It is only occasionally that all the terms from the term space need to be posed as queries to retrieve the entire set of documents from the database.

To find the order in which these candidate terms be chosen for generating the optimal query set $Q_{opt}$ , we model every document of the database in vector space model (VSM) which contains several dimensions representing the weight of each term. The conventional VSM has been modified in our approach by assigning weights according to the term's frequency of appearance at different positions in different document of the collection. **[Figure-4]** shows the index to be maintained for various statistics needed by our proposed approach.

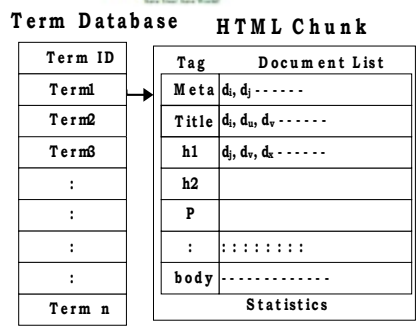

**Fig: 4. Term Statistics Index**

This paper presents a new term weighing scheme called VDF an acronym for variable document frequency that exploits the tag structure of the HTML language to deduce the magnitude of the term based on its position with respect to other positions and terms, where the terms have been extracted from the Metadata and the text associated with the documents in the collection.

Extracting the Metadata consists of pulling the terms (words) out of the text that wraps the search form in the interface and the text in the documents [10]. Specifically, it includes the terms of the *<p>* paragraph tag, the header tags *<h1>*, *<h2>* etc., the terms of the alt attribute in case of images

**[Figure-5]** shows the various locations for the metadata that are possible in the

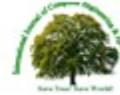

different HTML documents considered by our proposed model.

[Figure-6] below were retrieved by the system

```
D1: The Hidden Web crawler
D3: Crawling the Hidden Web
D6: Downloading the Hidden Web conten
```

**Fig: 6. Article Title Containing the Term "WEB"**

| Meta Data Locations |
| --- |
| **<title>** term₁ term₂ term₃……………………<\title> |
| **<meta** term₁ term₂ term₃……………………> |
| **<h1 >** term₁ term₂ term₃…………………term₆</h1> |
| **<p>** term₁ term₂ term₃………………………… |
| |
| ** |
| |
| ……………………term₅ </p> |

**Fig: 5. Meta Data Locations**

The weighing scheme has been devised based on the observation that the privileged location contains the Keywords. Thus, the term which occurs in the titles of some documents is likely to appear somewhere in other documents of the database."

For example: Consider a database that contains the documents regarding the research done by scientists in the area of "World Wide Web and Information retrieval". Many of the documents in the database have titles comprising of the term Web. So, when the term 'Web' is posed as query at the search box of the interface , some of the articles titled as shown in the

Also, all these documents contained the term hidden in the title position (along with various other positions) of the retrieved documents. It is also obvious that there will be many other documents that are likely to contain the term Hidden somewhere in the document (especially those that deal with web databases), though might not in the title of the documents. The following table in [Figure-7] shows some of the exemplary documents with the titles underlined and a part of the document where the term 'Hidden' appears.

Querying text databases for efficient Information Extraction

database. Hence *QXtract* will help deploy existing information extraction systems at a larger scale and for a wider range of applications than previously possible. In addition to improving efficiency of extraction, our automatic querying techniques can be used to extract a target relation from an arbitrary "hidden-web" database accessible only via a search interface [2]. As another example, our techniques could be

Modeling Query access to text databases

*Task 2: Text Database Summary Construction:* Many valuable text databases on the web have non-crawlable contents that are "hidden" behind search interfaces. Metasearchers are helpful tools for searching over many such databases at once through a unified query interface. A critical task for a metasearcher to process a query efficiently and effectively is the selection of the most promis-

Query planning for Deep Web databases

A recent and emerging trend in data dissemination involves online databases that are hidden behind query forms, thus forming what is referred to as the *deep web* [13]. As compared to the surface web, where the HTML pages are static and data is stored as document files, deep web data is stored in databases. *Dynamic* HTML pages are generated only after a user submits a query by filling an online form.

Building a database of online databases

The Web has been rapidly "deepened" by massive databases online. Recent surveys show that while the surface Web has linked billions of static HTML pages, a far more significant amount of information is "hidden" in the deep Web, behind the *query forms* of searchable databases. With its myriad databases and hidden content, this deep Web is an important frontier for information

**Fig: 7. Documents with the term 'Hidden' in paragraphs but not in title.**

Thus, the terms that always appear in the <p> tags of all the documents of the corpus, should be weighed in a normal way. However, the terms that appear in the <title> tags in some documents and in <p> tags in other should be considerable assigned weights. Likewise, the terms that always appear in only the <title> tag of all the documents are assigned the highest weight as they have the greatest semantic significance to describe the whole corpus of documents. Basically, it has been derived from the classical term frequency and Document frequency measures so as to assign an extra bonus weights to terms that are located at special positions in addition to the number of appearances within and among a documents.

The VDF scheme formally defines the weight of any term $q_i$ as:

$$
\begin{aligned}
W(q_i) &= Vardf(q_i) \\
&= \sum_{\substack{\text{for all positions} \\ \text{where} \\ \text{the term appears}}} varwt(pos) * df\,(pos) \\
&= varwt(titile)\ *\ df(title)\ +\ varwt(h1) \\
&\qquad\qquad * df\,(h1) \dots\dots\dots\dots.
\end{aligned}
$$

Where pos indicates the various positions occupied by the term in different documents, varwt(pos) is called the variable position weight and is a weight value associated with the indicated position pos ; df is a measure of the number of documents in which the term appears at the pos. The value of the variable position weight

$$
Varwt(pos) = \frac{Total\ number\ of\ documents}{No\ of\ total\ terms\ that\ occur\ in\ that\ positon}
$$

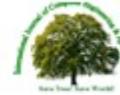

Now, suppose there are together of 10 distinguishing terms in the titles of the retrieved documents and the following data in **[Table-1]** has been extracted for the term 'Web'.

**[Table-1]**

| Term | Position | Documents |
|------|----------|-----------|
| Web | Title | D1,d3,d6 |
| | Heading | d6 |
| | Parah | D1,d3,d6 |

**Table: 1. Data Extracted from the Term 'Web'**

Then, for the term Web,

$$Varwt(title) = \frac{m}{10} \quad \text{And} \quad df(title)=3$$

Similarly, other variable weights and document frequencies can be calculated if a count of the total unique terms in each position from the retrieved documents is available. Thus, the variable document frequency of any term can be estimated.

## 4.2. The GREEDY COMPOSE Algorithm

When formulating the optimal set $Q_{opt}$, the crawl system need to measure the effectiveness of every candidate term $q_i$ includes the most effective term $q_{opt}$ and repeats the process until MAX terms are chosen where MAX is an empirical value as in users or crawler's specification which is always less than $|Q|$. The algorithm for choosing the set $Q_{opt}$, as per our approach has been termed as the GREEDY COMPOSE algorithm which is presented in figure 8 below:

The GREEDY COMPOSE Algorithm for generating $Q_{opt}$

Input - Keyword based search form Interface
- The query term space, Size MAX of $Q$
- Domain of the hidden database
Output - $Q_{opt}$

Method prepare an index of terms extracted from the various retrieved documents of the database
Associate each entry in the term statistics index with the document the position and the frequency of appearance of the term in the retrieved collection
Based on the statistics estimate the weight of each term and place the term having maximum weight at the highest position in the index or choosing as

1) Initially $Q_{opt} = \{domain\}$
2) Submit the form by filling the query and domain
3) Retrieve the result documents
4) Extract terms and Metadata
5) Vocabulary set=V{t_1, t_2, t_3,……t_n}, Q= Q + V - Q_{opt}
6) Create index by noting the position, frequency and the documents where the term appears

7) $Varwt(pos) = \frac{Total\ number\ of\ documents}{No\ of\ total\ terms\ that\ occur\ in\ that\ positon}$

8) $W(qi) = \sum_{for\ all\ positions} varwt(pos) * df$
    where
    the term appears

9) Rank the terms by their weights W extract the top ranked term

     $q_{opt}$= top ranked term
10) $Q_{opt} = Q_{opt} \cup \{q_{opt}\}$
11) Goto step 2 until Q is empty or $C(qopt)$ is not null
i.e. no more terms to be specified or no more new documents can be retrieved from the database

**Fig: 8. Greedy Compose Algorithm**

## [V] EXPERIMENTS

This section presents the results of experiments done for efficient term weighing and selection. The document database was related to research from the Information Retrieval community where initially 350 documents were randomly chosen for storage in the backend. Initially the term 'information' was posed as query which returned just 27 documents containing the term in one of the either positions. For each such query, an index structure was automatically created in MS-Access for recording the extracted terms along with the needed statistics. All the documents that were retrieved by posing a query were preprocessed to remove stop words and then broken into their constituent HTML tag structures so as to identify terms & their positions (by their bounding HTML tags) within the documents.

A successful optimal query is one that retrieves at least one unique document from the collection, in contrast to unsuccessful query that does not retrieve any useful documents. By unique or useful document we mean, a document that has not been retrieved by any of the previously issued queries.

Suppose if the number of queries successfully chosen as $Q_{opt}$ are S, the number of unsuccessful queries be U and the number of queries that did not return any results are denoted by N. The performance of our crawler has been measured by using the three most popular metrics:

• **Precision:** it is defined as the fraction of correctly issued queries to the total set of queries issued. Mathematically,

$$\text{Precision } P = S/(S+U)$$

• **Recall:** It is defined as the fraction of correctly issued queries to the actual set of correct queries

$$\text{Recall, } R = S/(S+N)$$

• **F-Measure:** It incorporates both precision and recall by making a tradeoff between the two. The general formula is

$F\alpha = \frac{(1+\alpha)P.R}{\alpha.P+R}$ $\alpha$ is the relative importance given to recall over precision. During our experiments we consider both recall and precision of equal importance and therefore set $\alpha=1$

$$\text{Thus, } F = \frac{2.P.R}{P+R}$$

The entire set of documents were retrieved by issuing just 42 queries, of which 32 queries brought up some new documents, 6 were not successful and 4 could not retrieve any documents.

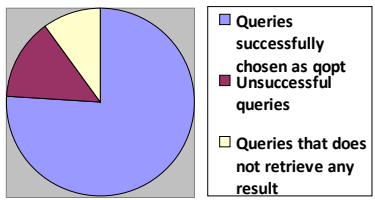

Fig: 9. Percentage depicting successful and unsuccessful queries

Thus

- %age of queries successfully chosen as $q_{opt}$ = 76

- %age of unsuccessful queries= 14

- %age queries that did not retrieve any results= 10

The values of the three parameters precision, recall and F-measure that are obtained by our system are 84%, 89% and 86.4%. **[Figure-10]** summarizes the results obtained by our proposed system.

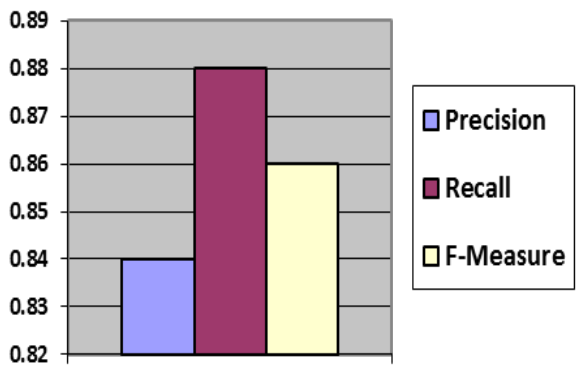

Fig: 10. Performance of the proposed system

## [VI] CONCLUSION

Choosing optimal query terms is the foundation of crawling the unstructured databases in the Hidden Web. Our paper proposes a new term weighing scheme that helps to solve the issues of optimal query selection by effectively selecting optimal terms. It gives a variable measure of the document frequency of terms because the location of the term which normally varies among the documents forms the basis for the assigned weights. The experiments conducted clearly shows that our approach efficiently crawls the Hidden Web pages due to the merit of our approach which lies in the various machine recognizable statistics derived from the skeleton of the document (HTML tags).

### REFERENCES

[1]   Michael Bergman, "The deep Web: surfacing hidden value". In the Journal Of Electronic Publishing 7(1) (2001).

[2]   S. Raghavan, H. Garcia-Molina. Crawling the Hidden Web. In: 27th International Conference on Very large databases (Rome, Italy, September 11-14, 2001) VLDB'01, 129-138, Morgan Kaufmann Publishers Inc., San Francisco, CA.

[3]    S. W. Liddle, D. W. Embley, D. T. Scott, S. H. Yau.   Extracting Data Behind Web Forms. In: 28th VLDB Conference2002 , HongKong, China.

[4]    L. Barbosa, J. Freire : Siphoning hidden-web data through keyword-based interfaces. In: SBBD, 2004, Brasilia, Brazil, pp. 309-321.

[5]    A. Ntoulas, P. Zerfos, J.Cho.  Downloading Textual Hidden Web Content Through Keyword  Queries. In: 5th ACM/IEEE Joint Conference on Digital Libraries (Denver, USA, Jun 2005) JCDL05, pp. 100-109.

[6]    E.Agichtein, L. Gravano. " Querying text Databases  for  Efficicnt  Information Extraction". In proceedings of the 19th IEEE International conference on Data Engineering (ICDE 2003) 2003

[7]    Z. Wu, Lu Jiang, Q. Zheng, J.Liu, "Learning , to surface Deep Web content". In proceedings of 24th AAAI conference on Artificial Intelligence, AAAI-10

[8]    E.Agichtein P. Ipeirotis,   L. Gravano. "Modeling   Query based access to Text Databases ". In the International Workshop on the Web and Databases (WebDB-2003)June 2003, San Diego, California.

[9]    Sonali Gupta, Komal Kumar Bhatia. "HiCrawl: A Hidden Web crawler for Medical Domain" in IEEE International Symposium on Computing and Business Intelligence,  ISCBI August-2013, organized by Cambridge Institute of Technology held at Delhi, India.

[10]   Sonali Gupta, Komal Kumar Bhatia: A system's   approach   towards   Domain Identification of Web pages, in Second IEEE international conference on Parallel, distributed and Grid computing , dec 06-08, 2012 organized by Jaypee University of Information Technology,  Shimla, H.P India.

[11]   Sonali Gupta, Komal Kumar Bhatia: Deep Questions in the 'Deep or Hidden'Web. In International Conference on Soft Computing for Problem Solving SocPros-Dec 28-30 , 2012, Rajasthan,   India proceedings by springer.